# Composite fabrics of conformal MoS$_2$ grown on CNT fibers: tough battery anodes without metals or binders

*Moumita Rana,[a] Nicola Boaretto,[a,b] Anastasiia Mikhalchan,[a] Maria Vila Santos,[a,c] Rebeca Marcilla,[d]\* Juan Jose Vilatela[a]\*\**

a. IMDEA Materials, Eric Kandel 2, 28906 Getafe, Madrid, Spain.

b. Centre for Cooperative Research on Alternative Energies (CIC energiGUNE), Basque Research and Technology Alliance (BRTA), Alava Technology Park, Albert Einstein 48, 01510 Vitoria-Gasteiz, Spain.

c. Escuela Técnica Superior de Ingeniería de Telecomunicación (ETSIT), Universidad Rey Juan Carlos, C/Tulipán s/n, 28933 Madrid, Spain.

d. Electrochemical Processes Unit, IMDEA Energy, Avda. Ramón de la Sagra, 3, 28937 Móstoles, Madrid, Spain

\*rebeca.marcilla@imdea.org; \*\*juanjose.vilatela@imdea.org

ABSTRACT: In the quest to increase battery performance, nanostructuring battery electrodes gives access to architectures with electrical conductivity and solid-state diffusion regimes not accessible with traditional electrodes based on aggregated spherical microparticles, while often also contributing to the cyclability of otherwise unstable active materials. This work describes electrodes where active material and current collector are formed as a single nanostructured composite network, consisting of macroscopic fabrics of carbon nanotube fibers covered with conformal $MoS_2$ grown preferentially aligned over the graphitic layers, without metallic current collector or any conductive or polymeric additives. The composite fabric of $CNTF/MoS_2$ retain high toughness and show out-of-plane electrical conductivity as high as 1.2 S/m, above the threshold to avoid electrical transport-limited performance of electrodes (1 S/m), and above control nanocomposite LIB electrodes (0.1 S/m) produced from dispersed nanocarbons. Cycled against Li, they show specific capacity as high as 0.7 Ah/g along with appreciable rate capability and cycling stability in low (108% capacity retention after 50 cycles at 0.1 A/g) as well as high current density (89 % capacity retention after 250 cycles at 1 A/g). The composite fabrics are flexible, with high tensile toughness up to 0.7 J/g, over two orders of magnitude higher than conventional electrodes or regular $MoS_2$ material, and full-electrode capacity above state of the art at different current densities.

**Introduction**

The quest to increase battery performance has sparked interest in new electrode architectures that are far from the conventional configuration of traditional electrodes essentially consisting of aggregated microparticles of active material and percolated conductive additives, glued together by a polymeric binder and supported on a metal foil. Non-conventional electrode/battery architectures can eliminate the need for polymeric binders[1] and enable dry processing,[2] in line with efforts to make manufacturing more sustainable. Furthermore, the use of high-aspect ratio constituents in battery electrodes confers them with mechanically augmented properties. These include self-standing electrodes with flexibility in bending through internal reinforcement provided by a network of nanocarbons,[3,4] electrodes with high thickness (high areal capacity) and reduced weight from non-active materials, high tensile ductility by patterning stretchable structures,[5] electrodes with high tensile toughness by using nanocarbon current collectors,[6] and ultimately, electrodes with carbon fibres that carry load-bearing and energy storage functions simultaneously (termed structural batteries).[7–9] A common feature of most emerging electrode architectures is an electrode structure resembling a structural composite, with significant load transfer between constituents compared to traditional electrodes with granular constituents.

A particularly promising strategy has been to produce electrodes containing a network of conducting nanocarbons (graphene, carbon nanotubes (CNT),[10] which not only provide high (out-of-plane) electrical conductivity but also act as scaffold to support and stabilise active material, thus increasing cyclability of high-energy density anodes[1] and cathodes.[11] Nanostructuring battery electrodes can give access to electrical conductivity and solid-state diffusion regimes not accessible with traditional electrodes based on aggregated spherical microparticles, and thus to near-theoretical capacity values, as demonstrated for a variety of chemistries.[1]

One avenue to produce nanocomposite battery electrodes is from dispersions of active material and nanofillers, as successfully demonstrated for graphene and/or CNTs with various chemistries (e.g., Si, lithium titanate, transition metal oxides, $MoS_2$). An alternative method is to integrate the active material into a pre-formed nanocarbon network, such as a fabric of CNT fibres.[12,13] The active material can then be directly infused or grown through wet-chemical,[14] electrochemical[15] or gas-phase processes.[16] A common feature of these composite electrodes with built-in current collector is a low electrical resistance for charge extraction.[15,17] One benefit is that these fabrics have (mass-normalised) electrical conductivity approaching Cu and tensile mechanical properties in the high-performance range,[18,19] eliminating the need for current collectors and unlocking structural properties.[6] We have previously shown, for example, that $MnO_2$ grown on porous CNT fibre fabrics exhibits and a high rate performance and excellent stability as anode for LIB.[15]

In this study, we have developed flexible $CNTF/MoS_2$ composite electrodes with $MoS_2$ (2H phase) grown aligned and conformally around CNTs in the fabric by an electrodeposition process. This eliminates all binders, additives, or metallic current collectors. These materials exhibited an in plane and out of plane conductivity in the order of $10^3$ and $10^{-1}$ S/m respectively, high tensile toughness of 0.8 J/g and large ductility up to 10%. When used as anode for LIB in a half-cell configuration, the composite electrode exhibited specific capacity over 700 mAh/g with 108 and 89 % capacity retention after 50 cycles at 0.1 A/g and 250 cycles at 1 A/g, respectively. Capacity normalised by full electrode weight surpasses the conventional electrodes with carbon additive, binder, and current collector. Detailed electrochemical analysis reveals that even at lower scan rates the lithium storage mechanism relies both on the conversion and pseudocapacitive processes, and the contribution of the latter increases at higher scan rate. These composites have combined storage and structural properties above those reported in literature.

**Experimental**

**Chemicals:** Butanol, ferrocene, ammonium tertrathiomolybate (99.9%), Sodium sulphate (99.99%). All chemicals were purchased from Sigma and used without further purification.

**Synthesis of free-standing CNT fibers:** The CNT fibers were synthesized by a gas phase chemical vapour deposition process. Here butanol, ferrocene and thiophene were used as source of carbon, catalyst, and promoter, respectively. These precursors were introduced from the top of a vertical furnace, passed through a temperature of 1250 °C under hydrogen atmosphere. The CNT fiber veils were collected from the bottom of the furnace by winding around a spool of Teflon film. After 30 minutes of collection, the free standing CNTF fabrics were taken out and used for electrochemical functionalization.

**Synthesis of CNTF/MoS$_2$ composite:** Prior to the deposition of MoS$_2$ in aqueous condition , to improve the hydrophilicity of the CNT fibers first the pristine CNTF fabrics were electrochemical functionalized at 2.5 V in an aqueous solution of 0.1 M Na$_2$SO$_4$ in two electrode configuration against a platinum mesh.[20] The electrodeposition of MoS$_2$ on the functionalized CNTF was performed using three electrode system, where Ag/AgCl (3M) electrode and a Pt mesh were used as reference and counter electrodes, respectively. A solution of the precursor molybdate salt and sodium sulphate was prepared by maintaining concentration of 10 mM and 0.1 M, respectively in Mili-Q water. The electrodeposition was performed using chronoamperometric technique at a voltage of -1.8 V for 3 to15 minutes. After deposition, the samples were thoroughly washed with water and ethanol, and dried under ambient condition. These samples were further heated at 600 °C in a horizontal furnace under continuous flow of argon for 1 hour with a heating ramp of 5 °C/min. The free-standing samples were employed for further characterizations. Unless stated otherwise, all composite samples presented in the manuscript were annealed.

**Physico-chemical and morphological characterization:** The samples were characterized thoroughly using field effect scanning electron microscopy (FESEM, FEI Helios NanoLab 600i), transmission electron microscopy (TEM, Talos F200X FEG, 200 kV), Raman spectroscopy (Ranishaw, fitted with a 532 nm laser source), selected area electron diffraction (SAED), powder X-ray diffractometer (PXRD, Cu Kα radiation, Empyrean, PANalytical Instruments) and Brunauer–Emmett–Teller (BET) adsorption (Micromeritics, TriStar II Plus Version 3.01).

**Mechanical characterization:** Tensile tests were performed using the Dynamic Mechanical Analyzer machine (DMA 850, TA Instruments) equipped with 18 N load cell. For this analysis, the tensile clamps designed for uniaxial deformation were used and tensile tests were performed in the force-controlled mode at the rate of 0.1 N min$^{-1}$ ramp, which corresponds to the strain rate of 0.1 mm min$^{-1}$. The CNTF/MoS$_2$ specimens were cut by sharp paper knife into rectangular strips with the width of 3 mm and the initial gauge length of 20 mm. The load-displacement curves were recorded, and the absolute and specific values of tensile strength, elastic modulus, and elongation to break were calculated afterwards. To calculate the apparent volumetric density, the mass of each sample was measured using high precision microbalance; and the thickness of samples was measured with a high precision digital micrometer. Specific properties, determined from knowledge of the load and linear density, are used to eliminate uncertainty with determination of cross sections and enable direct comparison of tensile properties for samples with different porosity.

**Electrical characterization:** In–plane electrical resistance was measured using the four probes technique to avoid contact contribution to resistance using a 2450 Keithley source–measurement unit. Strip electrodes were employed, spaced by 0.5 cm (the electrode width is 0.4 cm). To prove the homogeneity of the proved samples, different zones were tested. The electrical conductivity was further calculated by using the geometric parameters of the samples. Besides, the out – of plane electrical resistance was obtained using a two-probe method.

**Electrochemical characterization:** The electrochemical tests were performed using biologic electrochemical workstation (VMP300) and a Neware battery tester (CT-4008-5V10mA-164). The CNTF/$MoS_2$ composite samples were used as working electrodes in coin cell configuration (CR2032), where a piece of lithium metal foil (Sigma Aldrich) was used as negative electrode. $LiPF_6$ (1M) and 10% Fluoroethylene carbonate (FEC) in ethylene carbonate/diethyl carbonate (1:1 v/v, Solvionic) was used as electrolyte, and Whatman GF/D discs, along with Celgard 2400, facing the CNTF/ $MoS_2$ electrodes, discs were used as separators. All the cells were assembled in glove box, under argon atmosphere, with oxygen and moisture levels below 0.5 ppm. The cells were characterized using cyclic voltammograms (scan rates of 0.1, 0.25, 0.5, 1, 2.5, 5, 10, 25, 50, 100 mVs$^{-1}$, voltage window: 0-3V), and galvanostatic cycling (current densities of 25, 50, 100, 250, 500, 1000, 2500 and 5000 mA/g, voltage range: 3 to 0.01V). All voltages, mentioned throughout this manuscript, are with respect to Li/Li$^+$ redox potential. The specific capacity values of the CNTF/$MoS_2$ composites are normalized with respect to the total electrode mass.

**Results and discussion**

The composite electrodes were fabricated by chronoamperometric electrodeposition of MoS$_2$ directly on the CNT fibre (CNTF) fabrics. Ensuring growth of crystalline MoS$_2$ required mild functionalisation of the CNTF before electrodeposition,[21] and annealing after deposition.

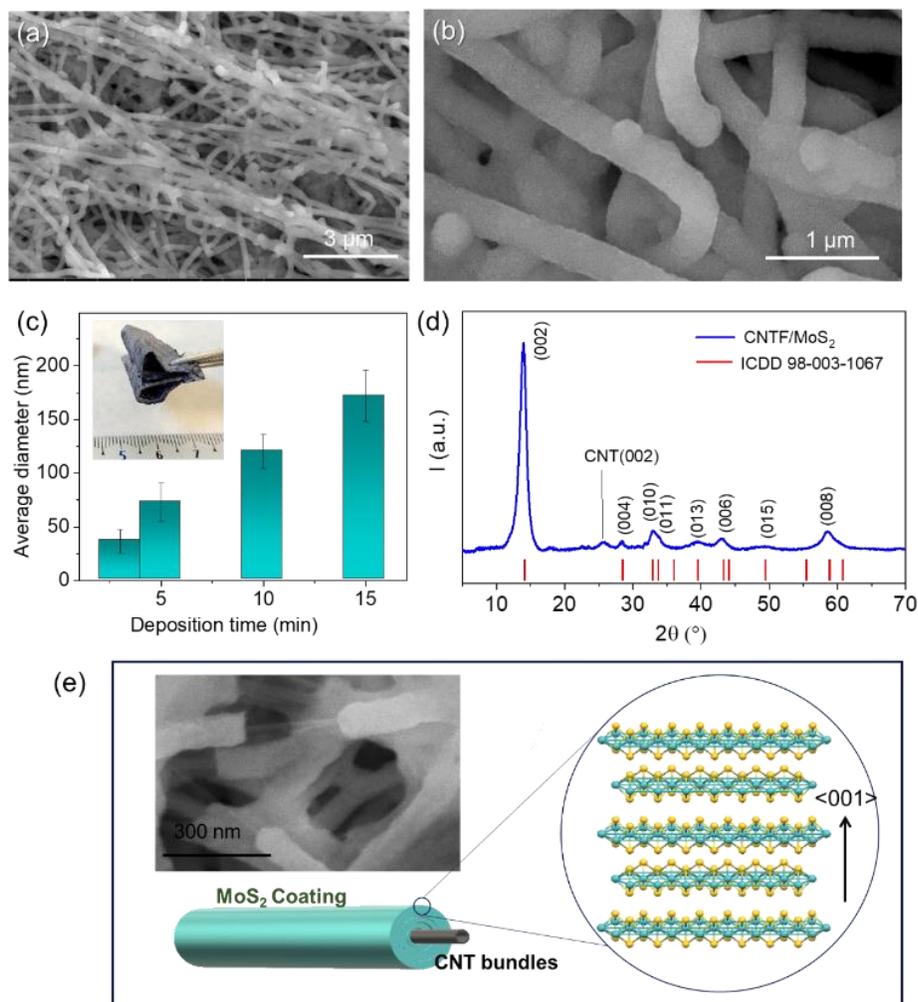

*Figure 1.* *(a, b) FESEM images of the CNTF/MoS$_2$ composite. (c) A bar diagram showing the dependence of the average diameter of the CNTF/MoS$_2$ core shell nanowires with electrodeposition time. (d) PXRD pattern of CNTF/MoS$_2$-84 composite. (e) Schematic of the CNTF/MoS$_2$ composite structure showing a conformal coating of MoS$_2$ around the CNTF bundles, where the layers of MoS$_2$ are oriented in the <001> direction. The inset image shows a broken part of the composite.*

After annealing, the macroscopic samples appear as shiny and highly flexible fabric, as shown in the inset in **Figure 1c** and **Figure S1.** The FESEM images of the samples are shown in **Figure 1 a, b, Figure S2 and Figure S3**. The growth pattern of $MoS_2$ follows the one-dimensional morphology of the pristine CNT fiber, leading to a uniform coating around the CNT bundles. Growth of $MoS_2$ does not disrupt the fabric structure, thus the material can be visualised as a percolated CNT network with a continuous conformal coating of the active material throughout the fabric. We measured the diameter of the core CNT bundles to be 7-20 nm, and the $MoS_2$ coating thickness increased with electrodeposition time (**Figure 1c** and **Figures S2-3**). The average thickness of the $MoS_2$ shell coated around CNT bundles are estimated to be 12, 30.5, 41.6 and 80 nm for the electrodeposition time of 3, 5, 10 and 15 minutes, respectively. The dependence of the electrode thickness, mass loading, average diameter, coating thickness and mass fraction of $MoS_2$ over electrodeposition time is summarized in the **Table 1**. For clarity, the samples with '*x*' mass fraction of $MoS_2$ are labelled as CNTF/$MoS_2$-*x* in the manuscript.

*Table 1. Morphological parameters of CNTF/$MoS_2$ composites produced with different electrodeposition times.*

| Sample (electrodeposition time) | Electrode thickness (average, μm) | Mass/Area (mg/cm$^2$) | Average bundle diameter (nm) | Average $MoS_2$ coating thickness (nm) | Mass fraction of $MoS_2$ (%) |
|---|---|---|---|---|---|
| Pristine CNT | 4 | 0.41 | ~16 [22] | - | 0 |
| CNTF/$MoS_2$ (5 minutes) | 28 | 1.33 | 72.9 | 30.5 | 69 |
| CNTF/$MoS_2$ (10 minutes) | 45 | 1.82 | 120.3 | 41.6 | 77 |
| CNTF/$MoS_2$ (15 minutes) | 61 | 2.57 | 172.1 | 80 | 84 |

The PXRD patterns of the CNTF/MoS$_2$ are shown in **Figure 1d** and **Figure S4** in SI. The small peak at 26° can be attributed to the (002) reflection from the CNT fibers. The rest of the peaks can be indexed based on the hexagonal 2H phase of MoS$_2$ (ICDD 98-003-1067), where the unit cell contains 2 MoS$_2$ layers in the c-direction. Interestingly, the indexed pattern shows presence of only the (*00l*) and (*0kl*) planes, which indicates that the basal planes of the MoS$_2$ predominate. This infers that the conformal coating of MoS$_2$ might be crystallographically oriented around the CNT core, as shown in **Figure 1e**. The orientation of MoS$_2$ relative to the CNTs can be further discerned from high resolution TEM on these samples (**Figure 2a, b** and **S5**). As shown in **Figure S5**, the (002) planes of MoS$_2$ (d=0.63 nm) are lying almost parallel to the growth direction of the CNTF. This texture is common in MoS$_2$/CNT materials produced by solvothermal[23–25] and solid-state[26] deposition methods. In the few reports on the use of electrodeposition techniques to grow MoS$_2$ on CNT, a random orientation has been previously observed.[27]

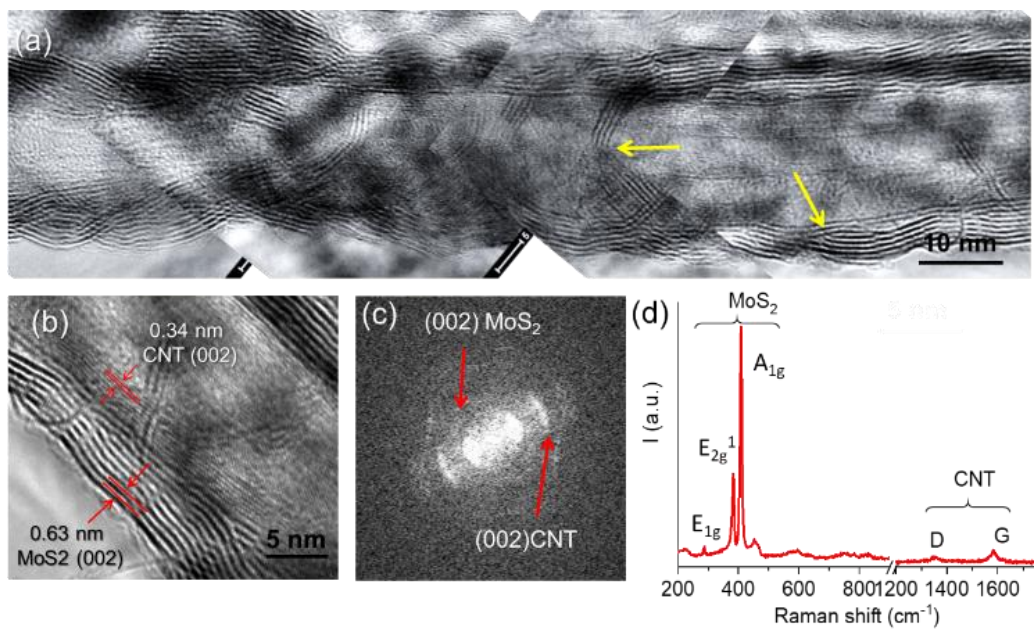

***Figure 2.*** *(a) Representative TEM image of CNTF/MoS$_2$ core-shell composite. The yellow arrows show presence of defects in the MoS$_2$ lattice. (b) HRTEM image of the composite showing the epitaxial growth of the MoS$_2$ (002) planes around the graphitic layers of CNT (002). (c) The FFT electron diffraction pattern of the composite. (d) Raman spectra of the CNTF/MoS$_2$ composite showing presence of low intensity D and G peaks from the CNT core.*

To investigate the MoS$_2$-CNT interface in detail, we performed high resolution TEM on a sample with thinner MoS$_2$ coating (3 minutes deposition). **Figure 2a** shows a representative HRTEM, where a CNT bundle, composed of 3 multiwalled CNT is uniformly coated with MoS$_2$, where the (002) planes of MoS$_2$ and CNT are almost aligned at the interface. The cylindrical symmetry and large thickness of the MoS$_2$ layers makes it difficult to determine if there is commensurability between CNT and MoS$_2$. Some commensurability is plausible though, given the evidence that small layers of MoS$_2$ readily align over the graphitic layers in crystallographic registry nearly free of strain.[28,29] However, MoS$_2$ layers show some degree of irregularity, particularly evidenced as basal plane edges analogous to edge dislocations, grain boundaries, undulated layers and other irregular domains formed due to the curvature of the CNTs and the inherently irregular bundle network structure.[30] The lattice mismatch between MoS$_2$ layers and graphitic surface of MWCNT (3.16 Å *vs.* 2.47 Å respectively), curvature of the MWCNTs and presence of kinks at the parallel joints of two nanotubes in a bundle can spontaneously generate strain in the MoS$_2$ lattice, which in turn leads to such defects.[30] Overall, as shown in **Figure 1e,** the composite fabric is essentially a continuous network with a core/ shell structure of CNTF/MoS$_2$ strongly bound by van der Waals forces.[31] The surface area of a representative CNTF/MoS$_2$ composite was estimated using BET isotherm analysis. In Figure S4b we show the adsorption and desorption profiles of N$_2$ at 77K temperature. The estimated surface area of the composite was 41 m$^2$/g for the sample with 69% MoS$_2$ mass fraction, with predominance of mesoscopic pores.

The presence of 2H phase of the MoS$_2$ shell was further confirmed by Raman spectroscopy. A representative Raman spectrum of the CNTF/MoS$_2$ composite is shown in the **Figure 2d.** The strong peaks around 400 cm$^{-1}$ can be attributed to vibrations from MoS$_2$. Two strong signature peaks at 408.5 and 382.1 cm$^{-1}$ can be assigned to the vertical (A$_{1g}$) and horizontal (E$_{2g}^1$) vibrations of the MoS$_2$ layers. The small peak at 286.7 cm$^{-1}$ can be assigned to the E$_{1g}$ vibrational mode, which further confirms presence 2H phase of MoS$_2$. The small peaks at 1583.6 and 1351.1 cm$^{-1}$ in **Figure 2d** are generated from the symmetric vibrations from graphitic (G) and defect (D) rich regions of the CNTF respectively. The intensity ratio of the D/G peaks was ~0.5.

**Mechanical and electrical properties of the composite electrodes:** Next, we determined basic mechanical and electrical properties, particularly to confirm the hypothesis that the electrode behaves as a composite of continuous CNTF fabric in a MoS$_2$ matrix. The mechanical behaviour of the composites was evaluated by uniaxial tensile test. **Figure 3a** shows the representative stress-strain curves of the CNTF/MoS$_2$ composites; and their specific tensile strength, modulus, and toughness (fracture energy) are summarized in **Table 2**. For comparison, we include reported values for MoS$_2$ networks and filler-type CNTF/MoS$_2$ composites, both of which are orders of magnitude below. Considering MoS$_2$ as a matrix in the CNTF/MoS$_2$ composite, the large increases in modulus, strength and toughness show effective reinforcement by the CNTF.

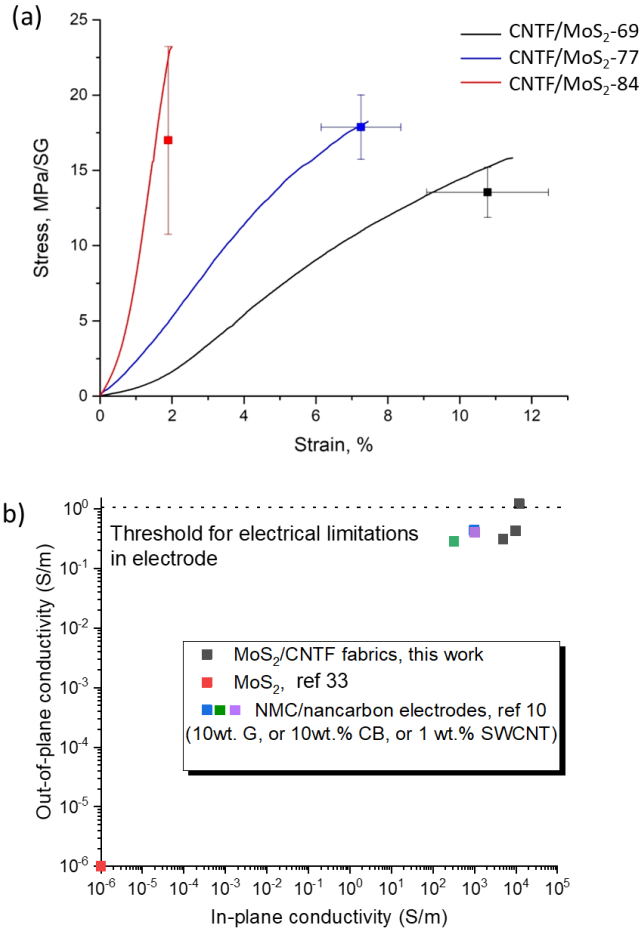

*Figure 3.* *(a) Stress –strain curves of the CNTF/MoS$_2$ composites. (b) Comparison of electrical conductivity of the CNTF/MoS$_2$ composites with pure MoS$_2$ and other electrodes (reference 10,33).*

*Table 2*: *Summary of mechanical and electrical properties of the CNTF/MoS$_2$ composites.*

| MoS$_2$ mass fraction | Specific strength, MPa/SG | Specific modulus, MPa/SG | Fracture energy, J/g | In-plane electrical conductivity (S/m) | Out-of-plane electrical conductivity (S/m) |
|---|---|---|---|---|---|
| 69% | 14.5 ± 1.7 | 164 ± 24 | 0.70 ± 0.19 | 1.19 (± 0.086) x 10$^4$ | 1.22 ± 0.03 |
| 77% | 18.0 ± 2.1 | 324 ± 52 | 0.67 ± 0.13 | 9.80 (± 0.29) x 10$^3$ | 0.43 ± 0.02 |
| 84% | 17.0 ± 6.2 | 1245 ± 365 | 0.13 ± 0.06 | 4.89 (± 0.29) x 10$^3$ | 0.31 ± 0.02 |
| 100%[32,33] | 0.1 | 1 | - | 1x10$^{-6}$ | 1x10$^{-6}$ |

However, from **Figure 3a** it is evident that with increasing thickness of the MoS$_2$ coating layer around the CNT bundles, the average specific strength of the samples remains almost same, whereas the samples lose their ductility and become more brittle in nature, the strain-to-break decreased from 10.4% for the CNTF/MoS$_2$-69 to 1.9% for CNTF/MoS$_2$-84. Moreover, we note that the modulus of these composite follows neither a corrected rule of mixtures[34] nor percolation relations observed in related systems (**Figure S6**).[32] We hypothesise that the results exhibit the transition from a porous, network-like composite to a composite with a continuous matrix. Tensile deformation is governed by the network structure of the CNTF; as the MoS$_2$ coating increases it produces stiffening by blocking fabric deformation mechanisms as internal fabric pores are occupied by MoS$_2$, such as rotation and transverse compaction of the network present in the pure fabric. This mechanism would explain the increase in stiffness for higher MoS$_2$ mass fraction, at nearly constant strength.

We tested the in-plane and out-of-plane electrical conductivity of the self-standing CNTF/MoS$_2$ composite fabrics. Out-of-plane conductivity values for all composite samples are above 0.3 S/m, close to or above the minimum to ensure electrode performance is not limited by electrical transport of ≈1 S/m observed in nanocomposite cathodes, and above limiting values obtained with cathode nanocomposites produced from controlled dispersions with different nanocarbons: graphene (>12 wt.%), carbon black (>12 wt.%) or SWCNTs (<1.3 wt.%).[10] The electrical conductivity of these samples is entirely dominated by the percolated network of the constituent CNT fibre, which remains largely unaltered through the incorporation of the MoS$_2$ phase. The MoS$_2$ layer is deposited on exposed surfaces of CNT bundles but does not disrupt the interfaces between adjacent CNTs responsible for charge transfer. Indeed, longitudinal conductivity for the different composite electrodes follows a dependence close to rule of mixtures; the presence of MoS$_2$ contributes to the composite density but not significantly to the conduction mechanism. The samples are very anisotropic though, with a ratio of in-plane to out-of-plane conductivity of 10$^4$, reminiscent of the high degree of alignment of CNTs in the plane, and significantly above 10$^2$-10$^3$ observed in electrodes produced from dispersions of shorter CNTs.[10] These results highlight important differences of the electrode composite route described in the manuscript compared to wet-chemical methods.

**Electrochemical properties:** In the study of CNTF/MoS$_2$ composites as anodes for lithium batteries, the first aspect of interest is to identify the different energy storage processes in the electrode, both from MoS$_2$ and from the CNTs. **Figure 4a** shows the cyclic voltammograms of CNTF/MoS$_2$//Li cells at a scan rate of 0.1 mV s$^{-1}$ (first, second and fifth cycles, for the anode with 69 wt.% MoS$_2$). Even though the features of the consecutive anodic cycles are quite similar, distinct difference is observed in the cathodic cycles, which are very typical for MoS$_2$.[35,36] During the first lithiation (black curve), the cathodic peak at 1.0 V can be attributed to the insertion of lithium in the layered 2H phase of the MoS$_2$, which leads to the metallic 1T phase with composition of Li$_x$MoS$_2$.[37,38] The second peak around 0.5 V can be related to the conversion of this lithiated 1T phase to Mo and LiS$_2$ along with formation of solid electrolyte interphase (SEI). The deep lithiation peak could be due to the previously mentioned conversion reaction, together with the lithiation of the CNTF scaffold.[39] In the following delithiation cycle, the strong anodic peak at 2.3 V is originated from the oxidation of Li$_2$S to S$_8$, whereas two small humps at 1.2 V and 1.8 V can be related to delithiation of residual Li$_x$MoS$_2$ or to the formation of some MoS$_2$,[11,35] with possibly also some contribution from hysteretic delithiation of CNTF.

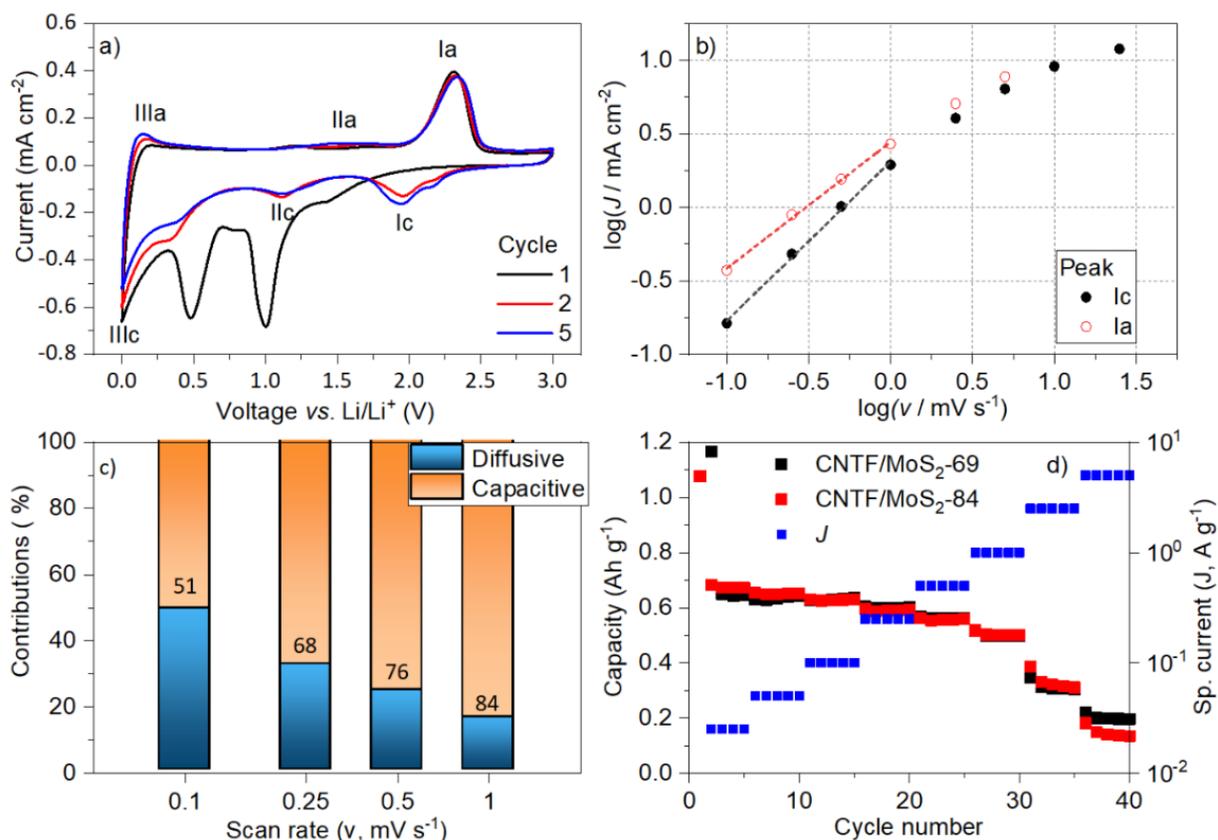

***Figure 4.*** *(a) Cyclic voltammograms of CNTF/MoS$_2$-69 at a scan rate of 0.1 mV s$^{-1}$ for the first, second and fifth cycles. (b) Log-log plot of the intensity of peaks Ic and IIIa versus the scan rate during CVs. (c) Variation of deconvoluted contributions of the diffusion controlled and pseudocapacitive process at different scan rates. (d) Specific capacities of CNTF/MoS$_2$-69//Li and CNTF/MoS$_2$-84//Li cells, during lithiation, at different specific currents. Specific capacities are normalised by the mass of the whole electrode.*

During the second cathodic scan (red curve in **Figure 4a**), the peak at 1.9 V (Ic) corresponds to the reduction of sulphur to Li$_2$S.[35] The peak IIc at 1.0 V can be related to the lithiation of MoS$_2$, reformed during the previous anodic scan. A shoulder at ~0.35 V could be attributed to the occurrence of conversion reaction, which would be in accordance with the formation of Li$_x$MoS$_2$ at 1.0 V. The shape of the CV reproduces thereafter, confirming the chemical stability of the composite electrodes. From the second cycle onwards, a new weak anodic peak appears at about 0.2 V, which is originated from the non-hysteretic delithiation of the CNTF scaffold.[39] The two anodic peaks at 1.2 V and 1.8 V merge in a broad peak centred at 1.7 V (IIa in **Figure 4a**).

Due to the large surface specific area of the composite and presence of defect sites, charge storage in CNTF/MoS$_2$ is expected to occur partially as a surface process and, therefore, it may be characterized by capacitive-like kinetics. To deconvolute the contributions from diffusion-limited and capacitive-like processes, we performed CVs at scan rates from 0.1 to 100 mVs$^{-1}$ (**Figure S7**). The logarithm of the peak current of the most intense peaks (Ia and Ic) increases linearly with log($v$) up to scan rates of 1 mV s$^{-1}$ (**Figure S7a**). Below this threshold, the slope of the log-log plot of the peak current log($J$) versus the scan rate log($v$) is 1.1 for peak Ic and 0.85 for peak Ia (**Figure 4b**). A slope of 1 is expected for capacitive processes, in which the current is proportional to the scan rate ($J \propto v$), and a slope of 0.5 for diffusion-controlled processes, in which the current is proportional to the square root of the scan rate ($J \propto v^{0.5}$). The results thus indicate that the processes underlying the two peaks possess a mixed electrochemical response but tending to be more capacitive type. A more detailed kinetic analysis was carried out, in the scan-rate range between 0.1 and 1 mV s$^{-1}$, by employing the following equation: [40]

$$J(V) = K_1 v + K_2 v^{1/2} \qquad (eq.1)$$

Here $v$ is the scan rate, $K_1$ and $K_2$ are scan rate independent constants. $K_1 v$ and $K_2 v^{1/2}$ stand for the capacitive and diffusive contribution respectively (e.g. **Figure S7b, c**). **Figure S7d-g** show the deconvoluted CVs of CNTF/MoS$_2$-69 at scan rates between 0.1 and 1 mV s$^{-1}$, where the blue regions stand for the diffusion-controlled process. At a scan rate of 0.1 mV s$^{-1}$, the processes underlying the main delithiation peaks (Ia and IIIa) are mostly diffusion-controlled, whereas the main lithiation processes (Ic and IIIc) show mixed capacitive/diffusion control. In the **Figure**

**S7d-g,** the blue regions at >0.2V can be related to the diffusion-controlled conversion of the MoS$_2$ layer, whereas the same at shallow potentials for all scan rates, indicates a significant contribution from the CNT fibers in the composite. In the other regions, the process is mainly capacitive. At higher scan rates the capacitive contribution obviously increases, due to the stronger dependence with the scan rate. The capacitive contribution to the total specific capacity at a scan rate of 0.1 mV s$^{-1}$ was estimated to be 51 %, which increases up to 84 % at a scan rate of 1 mV s$^{-1}$ (**Figure 4c**). This observation confirms that, at faster scan rates, the lithium storage mostly relies on the pseudocapacitive processes, owing to the large surface specific area as well as presence of defects in the composite.

The cell performance of the composites was further evaluated by performing galvanostatic charge-discharge in CNTF/MoS$_2$//Li half-cells. The potential-capacity profiles of the first two cycles at a specific current of 0.025 A g$^{-1}$ for CNTF/MoS$_2$//Li cells are shown in **Figure S8a**. During the first lithiation, the initial capacities were found to be ca. 1.2 Ah g$^{-1}$, with reversible capacities of 0.65 Ah g$^{-1}$. Note that this capacity is normalised by the mass of the whole electrode, including the CNT built-in current collector. Most of the irreversible capacity during the first lithiation could be originated from the formation of the lithiated 1T phase (pseudo-plateau between 1.5 and 1.0 V) and the SEI formation (plateau at 0.6 V). After the first lithiation, the potential profile shows pseudo-plateaus between 2.0 and 2.5 V and, during lithiation, below 0.5 V, concordantly to the CVs. From the second cycle onwards, the voltage profiles are reproduced completely, further confirming excellent electrochemical stability of the material. No appreciable difference was observed for anodes with 69 or 84 wt.% MoS$_2$ (**Figure S8**), which is due to the relatively small difference in mass fraction and the fact that the CNTF phase contributes to the lithiation process.

**Figure 4d** shows the specific capacities of CNTF/MoS$_2$ at specific currents ranging from 0.025 to 5 A g$^{-1}$ (~ from C/27 to 7.5 C) (for 69 and 84 wt.% of MoS$_2$ and normalised by whole electrode mass). The corresponding potential profiles at different specific currents for MoS$_2$/CNTF-69//Li are reported in **Figure S8b**. The composite shows excellent rate capability, which is in accordance with the large capacitive contribution observed in the CVs and its high electrical conductivity. Interestingly, the two samples, with different content of MoS$_2$, show comparable rate capability (**Figure S8c**), except at the highest tested specific current (5 A g$^{-1}$). At this specific current, equivalent to a current density of 7 mA cm$^{-2}$, CNTF/MoS$_2$-69//Li cells deliver 0.2 Ah g$^{-1}$, i.e. with capacity retention of 32 %. At the same specific current, CNTF/MoS$_2$-84//Li cells deliver 0.13 Ah g$^{-1}$ (20 % capacity retention).

The rate capability of these composite electrodes was further assessed by means of equation 2:[41]

$$Q = Q_0 \left(1 - (Rt)^n \left(1 - e^{-(Rt)^{-n}}\right)\right) \quad \text{(eq. 2)}$$

where $Q$ is the specific capacity, $Q_0$ is the capacity at zero-rate, $R$ is the effective rate (i.e. the current-capacity ratio, $J/Q$), $t$ is a characteristic time constant and $n$ indicates the type of process limiting the electrochemical performance ($n = 0.5$ for diffusive control and $n = 1$ for resistive control). $Q_0$ is for both electrodes ca. 0.6 Ah g$^{-1}$, which is close to the experimental value at 0.025 A g$^{-1}$ (**Figure S8c**). The exponent $n$ is in both the cases 0.7, indicating a mixed diffusive/capacitive control process. The time constant ($t$) is calculated to be 220 s for CNTF/MoS$_2$-69 and 260 s for CNTF/MoS$_2$-84, concurrently with the lower capacity of the latter at high rates. The transport coefficient $\Theta$, defined as $L^2/t$, where $L$ is the thickness of the electrode, is thus 3.56 x10$^{-12}$ m$^2$s$^{-1}$ for CNTF/MoS$_2$-69 and 14.31 x10$^{-12}$ m$^2$s$^{-1}$ for CNTF/MoS$_2$-

84. The transport coefficient has been shown to scale inversely with the volumetric capacity.[41] Given a volumetric capacity of ca. 300 and 150 mAh cm$^{-3}$ for CNTF/MoS$_2$-69 and CNTF/MoS$_2$-84, respectively, the calculated transport coefficient falls within the expected range.

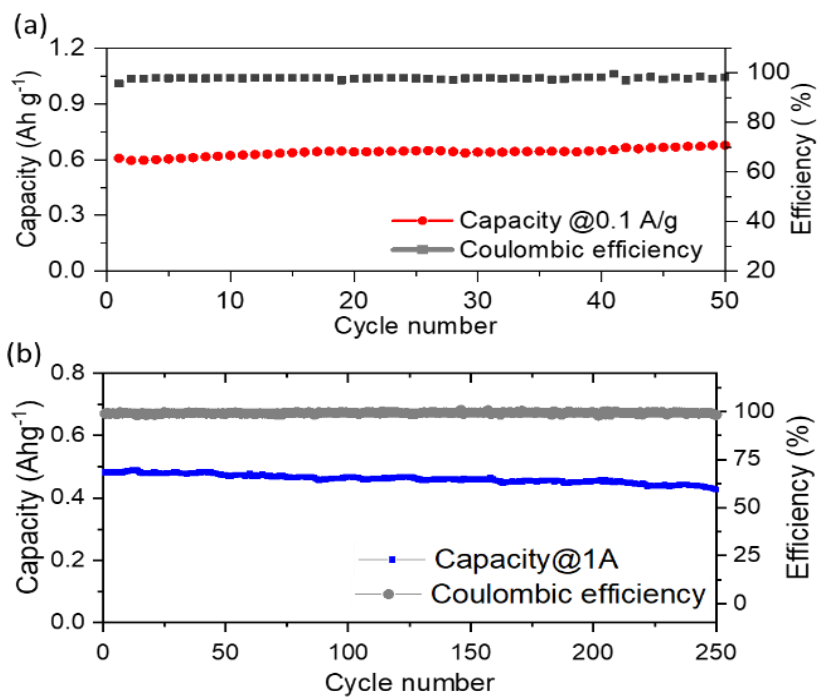

**Figure 5.** *Galvanostatic cycling of the CNTF/MoS$_2$-69 composite at a) 0.1 A g$^{-1}$ (~C/6.7) and b) at 1 A g$^{-1}$(~1.5 C) in a voltage window of 3-0.01V vs Li/Li$^+$.*

The cyclability of CNTF/MoS$_2$ composite was evaluated under galvanostatic condition. **Figure 5a and b** show the stability profile of the CNTF/MoS$_2$ composite cells at 0.1 and 1 A g$^{-1}$, respectively. At 0.1 A g$^{-1}$, a slight capacity increase is observed, from 0.6 Ah g$^{-1}$ to 0.65 Ah g$^{-1}$, with average coulombic efficiency of 97.8 %. The increase in specific capacity is attributed to an electrochemical milling process slightly increasing specific surface area of the composite over several cycles, and to an increased capacity contribution of the CNTF scaffold. The effect can be followed through the evolution of the potential profiles between cycles 1 and 50 (**Figure S8d**).

The capacity increase is accompanied by the appearance of a pseudo-plateau in the intermediate potential region (at about 1.2 V during lithiation, and 1.5 V during delithiation). Conversely, the capacity contribution of the high voltage plateau, which is strictly related to the electrochemical activity of $MoS_2$, remains constant. This suggests that the additional capacity may be related to an additional contribution of the CNTF scaffold, due to introduction of defects on the CNT fibers.[39] The effect is not observed at high current densities since in this case the lithium ions accumulation occurs mainly on the composite surface, and therefore the CNT cores are less affected by the lithiation process. As shown in the **Figure 5b**, the capacity retention after 250 cycles at 1 $Ag^{-1}$ current density was 88.6 %, with an average coulombic efficiency of 99.31%. Such electrochemical cyclability of these materials is related to the presence of the CNT network containing the $MoS_2$ material throughout cycling and despite some degree of electrochemical milling. **Figure S9** shows examples of CNTF/$MoS_2$ FESEM images after galvanostatic cycling experiments, confirming the stability of conformal composite structure even after extended electrochemical cycling.

**Comparison to other composite electrodes:** **Figure 6** presents a comparison of the properties of the composites studied in this work, with reports of related electrodes with a (nano)composite structure. In first instance, we plot capacity normalised by active material, against current density (**Figure 6a**). This normalisation is common in the literature but has limited significance when only $MoS_2$ is taken as active material, since it ignores the contribution to capacity from other materials in the composite electrodes. For electrodes with very high capacity (e.g. Si), the contribution from graphitic material may be negligible, but as shown in the present study, for $MoS_2$ that is not the case. Hence, for the materials in the present study, the active material includes the CNT network in the $MoS_2$/CNT composite electrode. From the graph in Figure 6a it is possible to identify the materials at the higher end as those in which $MoS_2$ consists of large, individualised monolayers of $MoS_2$, which are likely to have different storage mechanisms through a lower defect density and higher surface area. Next, we present the same data with capacity normalised by full electrode mass, including the current collector (**Figure 6b**). This comparison shows that the present $CNTF/MoS_2$ composites are above state-of-the art at nearly all current densities, highlighting the benefit of introducing the CNT current collector into the electrode material itself.

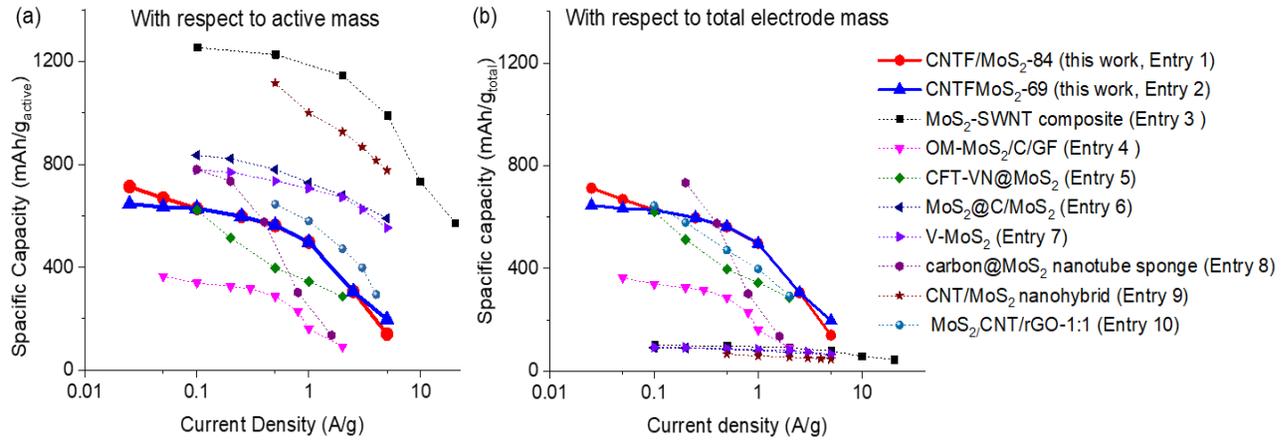

*Figure 6. A comparison plot of the specific capacity with respect to (a) active mass and (b) total mass of the electrode over current density of the electrodes. The details of the references are summarized in the table ST1. Note that for the composites in this work the active mass is considered including the CNTs.*

Finally, another aspect of interest of the present electrodes are their mechanical properties, which we present in **Figure 7** as a plot of specific capacity against tensile fracture energy. Although they have a modest strength, in the range of a soft metal like copper, they have high toughness. Their tensile fracture energy, which may be taken as indication of overall robustness for handling and further processing, is significantly above all reports on $MoS_2$-based electrodes and orders of magnitude above $MoS_2$ (~1 x$10^{-3}$ J $g^{-1}$).[32] Mechanically, the present composites are only below electrodes with structural carbon fibres[1] or with CNT fibre fabrics containing ceramic-like active material,[6] both with significantly lower capacity.

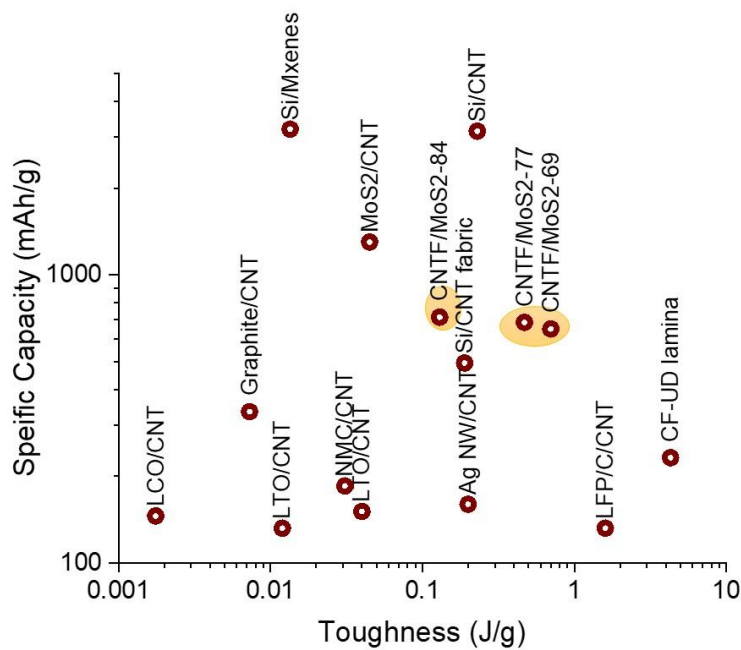

*Figure 7. A comparison plot of the specific capacity over corresponding toughness of the electrodes. The details of the references are summarized in the table ST2.*

**Conclusion**

In conclusion, we have demonstrated that electrodeposited CNTF/MoS$_2$ can be used as a tough anode for lithium-ion battery. This synthesis method offers uniform deposition of MoS$_2$ conformal around the CNT bundles forming a composite structure. The thickness of the MoS$_2$ shell can be modulated by varying the electrodeposition time of the MoS$_2$. Interestingly the orientation of MoS$_2$ layers around the graphitic CNT wall mostly maintains a near epitaxial relationship and are in close contact with the CNTs throughout the whole interface. Both in- and

out-of- plane electrical conductivities are high, above related nanocomposite electrodes produced by alternative routes based on aggregation of dispersed nanofillers. When used as Li-ion battery anode in a half cell configuration, the composites show specific capacity of ~0.7Ah/g with high-rate capability and electrochemical stability. A significant capacity is due to capacitive processes, as well as to Li storage in the CNTs. The composites show specific toughness of 0.3-0.8 J/g, which is superior to most of the high-capacity structural electrodes. The carbon content in the $CNTF/MoS_2$-74 composite is almost equivalent to the conventional recipe of electrode preparation, while avoiding the presence of polymeric binder and current collector, thereby gaining on the mass saving. This study establishes an alternative route to prepare current collector and binder free tough anode without compromising on their mechanical stability and specific capacity.

AUTHOR INFORMATION

**Corresponding Author**

*rebeca.marcilla@imdea.org; **juanjose.vilatela@imdea.org

**ACKNOWLEDGMENT**

The authors are grateful for generous financial support provided by the European Union Horizon 2020 Programme under grant agreement 678565 (ERC-STEM), by the MINECO-Spain (RyC-2014-1511, HYNANOSC RTI2018-099504-A-C22), by the Air Force Office of Scientific Research of the United States (NANOYARN FA9550-18-1-7016) and by "Comunidad de Madrid" FotoArt-CM project (S2018/NMT-4367). A.M. acknowledges funding from the European Union's Horizon 2020 research and innovation programme under the Marie Skłodowska-Curie grant agreement 797176 (ENERYARN). M.V. acknowledges the Madrid

Regional Government (program "Atracción de Talento Investigador", 2017-T2/IND-5568) for financial support.